# Extreme-Band-Gap Semiconductors with Shallow Dopants and Mobile Carriers

Sieun Chae[1,2,†], Nocona Sanders[1], Kelsey A. Mengle[1], Amanda Wang[1], Xiao Zhang[1], Jon Lafuente Bartolome[3,4,5], Kaifa Luo[3,4], Yen-Chun Huang[2], Feliciano Giustino[3,4], John T. Heron[1], and Emmanouil Kioupakis[1,*]

[1]Materials Science and Engineering, University of Michigan, Ann Arbor, MI 48109, United States

[2]School of Electrical Engineering and Computer Science, Oregon State University, Corvallis, OR 97331, United States

[3]Oden Institute for Computational Engineering and Sciences, The University of Texas at Austin, Austin, TX 78712, United States

[4]Department of Physics, The University of Texas at Austin, Austin, TX 78712, United States

[5]Department of Physics, University of the Basque Country, Leioa, 48940 Bizkaia, Spain

[*]Correspondence to: kioup@umich.edu

[†]Correspondence to: sieun.chae@oregonstate.edu

**The conventional distinction between semiconductors and insulators is often based on the magnitude of the band gap, with materials exhibiting gaps wider than 3 eV typically classified as insulators.[1] However, the emergence of ultra-wide-band-gap (UWBG) semiconductors such as AlGaN, diamond, BN, and $Ga_2O_3$ challenges this paradigm for materials classification[2] and raises fundamental questions about the upper bound of band gaps compatible with semiconducting behavior. Here we develop a computational-discovery**

**strategy to identify semiconductors with band gaps exceeding that of AlN (6.2 eV), while retaining essential semiconducting properties such as shallow dopants and mobile charge carriers. We discover that materials composed of light elements in densely packed crystal structures exhibit wide band gaps and light carrier effective masses that enable shallow dopants, high mobility, and weak polaron binding. By applying the hydrogenic Bohr model and first-principles defect calculations – validated against available experimental data – to screen for materials with shallow dopants, we identify dopable compounds with gaps as wide as 9.5 eV that nonetheless host mobile charge carriers. Our findings demonstrate that semiconducting behavior persists even at extreme band gaps, far beyond conventional upper bounds traditionally associated with semiconductor materials.**

Band theory classifies materials as metals, insulators, or semiconductors based on their electronic band structure and resulting electrical resistivity[3]. In semiconductors, electrical conduction arises from thermally excited carriers across the band gap, making the magnitude of the band gap a longstanding criterion for distinguishing semiconductors from insulators. Materials with relatively narrow gaps ($E_g$ < 3 eV) are typically considered semiconducting, while those with wider gaps are often identified as insulators[1]. However, semiconductors are defined not solely by their resistivity, but by their tunable conductivity through doping or electrostatic fields, an essential feature for modern microelectronic devices. Although materials with narrow gaps such as silicon (~1.1 eV) dominate current technology, wider-gap semiconductors are increasingly pursued for high-power electronics due to their higher dielectric breakdown field, which increases superlinearly with band gap.[4] Recently, ultra-wide-band-gap (UWBG) materials such as AlGaN, diamond, $\beta$-$Ga_2O_3$, and cubic BN (c-BN), have emerged as promising candidates for next-generation power electronics. Yet their band gaps, ranging from 3.5 eV to 6.2 eV, challenge the conventional gap-based classification of semiconducting materials and raise fundamental questions: How should semiconductors be defined? What material properties enable electrical conduction in UWBG compounds? And what is the upper limit of the band gap for a material to still exhibit semiconducting behavior?

Semiconductors are distinct from insulators by their ability to host delocalized mobile charge carriers (electrons and holes) that conduct electricity with high mobility, $\mu = \frac{e \cdot \tau}{m^*}$, where $\tau$ is the scattering time and $m^*$ is the carrier effective mass. Free carriers are typically introduced via intentional doping with impurity atoms. Within the hydrogenic Bohr model, the ionization energy of dopants is given by $E^{Bohr} = 13.6 \frac{m^*}{\varepsilon_0^2}$ (eV), where $\varepsilon_0$ is the static dielectric constant. For dopants to contribute free carriers at room temperature, their ionization energy must be comparable to or smaller than the thermal energy ($k_B T \approx 25$ meV). Free carriers can also arise at polar interfaces, where polarization discontinuities give rise to two-dimensional electron or hole gases.[5–7] For conduction to occur, these free carriers must remain delocalized. Yet, strong coupling to lattice distortions can localize them to form polarons, suppressing the mobility. The formation energy of a polaron, which acts as an additional barrier for hopping transport, is given in Hartree energy units ($E_{\text{Ha}} \simeq 27.2$ eV) by $\frac{E_{\text{polaron}}}{E_{\text{Ha}}} = -\frac{1}{3\pi} \frac{m^*/m_e}{\varepsilon_{\text{eff}}^2}$, where $\varepsilon_{\text{eff}}$ is the effective dielectric response obtained by separating the ionic ($\varepsilon_0$) from the electronic ($\varepsilon_\infty$) dielectric response, $\frac{1}{\varepsilon_{\text{eff}}} = \frac{1}{\varepsilon_\infty} - \frac{1}{\varepsilon_0}$.[8] Therefore, a light effective mass ($m^*$) emerges as a critical material parameter for semiconducting behavior, as it simultaneously enables shallow dopants, high carrier mobility, and low polaron dissociation energies. In general, however, carrier effective masses tend to increase with increasing band gap (Fig. 1e, f and Extended Data Fig. 1), resulting in deeper dopants and stronger localization in UWBG materials. Nonetheless, materials such as diamond[9] and c-BN[10] demonstrate efficient electrical conduction despite their ultra-wide gaps, which motivates the search for materials that similarly defy these conventional trade-offs. While recent high-throughput studies have derived guidelines based on chemistry for finding light carrier effective masses and thus efficient n- and p-type doping in oxides,[11,12] the identified compounds include heavy elements and exhibit only moderate band gaps. General design principles to identify materials that simultaneously combine ultra-wide band gaps with low carrier effective masses are still lacking, hindering the discovery of new UWBG semiconductors that can overcome the limitations of current materials.

In this work, we integrate chemical and crystal-structure criteria into a unified computational framework to identify new UWBG semiconductors and probe the maximum band gap compatible with semiconducting behavior. We establish simple chemistry–structure–orbital design principles that predict materials in which ultra-wide gaps can coexist with shallow ionization energies, weak polaron binding, and high carrier mobilities. Guided by these insights, we identify and validate several new UWBG semiconductors with band gaps as wide as 9.50 eV. These materials can achieve record-high dielectric breakdown fields, opening new opportunities in compact power electronics and resilient microelectronics for extreme environments.

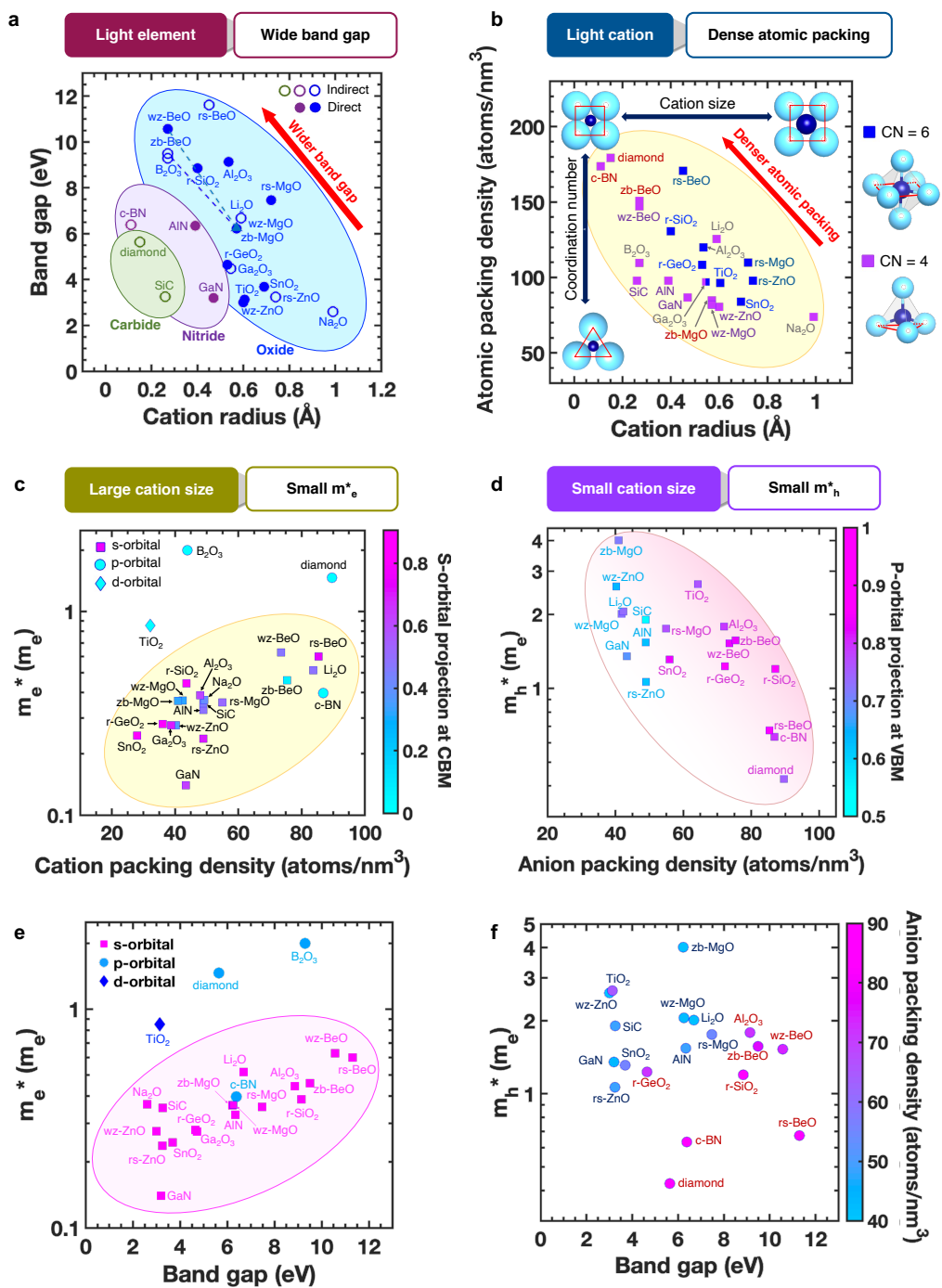

**Fig. 1. Workflow for screening extreme-band-gap materials with light electron and hole effective masses. a,** The band gaps of oxides, nitrides, and carbides are wider for lighter cations. **b,** The atomic packing is denser for materials with smaller cations or octahedral coordination

[coordination number (CN) is 6]. **c,** The electron effective mass is generally lighter for materials with larger cation *s* orbitals (sparser cation packing). **d,** Hole effective masses generally decrease with increasing anion packing or valence bands with mixed orbital character. **e,** The electron effective masses generally increase with increasing gap when having the s-orbital characteristics of conduction band. **f,** The hole effective masses generally increase with increasing gap when having the similar anion packing density.

Figure 1 identifies materials that combine ultra-wide band gaps with light carrier effective masses – key features of viable UWBG semiconductors. A general trend emerges from the atomic chemistry: lighter elements tend to form wider-gap compounds, as their lower principal quantum numbers yield larger energy differences between occupied and unoccupied atomic orbitals (Fig. 1a). This trend motivates a focus on carbides, nitrides, and oxides, which also happen to be well-known UWBG families. Among these, typically oxides exhibit the widest gaps, driven by the higher electronegativity of oxygen. Materials with gaps exceeding 7 eV are predominantly oxides of cations from the second and third row of the periodic table (e.g., $Be^{2+}$, $Mg^{2+}$, $B^{3+}$, $Al^{3+}$, and $Si^{4+}$).

We also examine the chemical and structural factors that control carrier effective masses. In binary compounds, the conduction band minimum is primarily derived from the lowest unoccupied cation orbital: the *s* orbital for main-group elements, the *p* for heavy *p*-block cations (e.g., $Pb^{2+}$ or $Bi^{3+}$), and the *d*/*f* orbitals for transition metals and rare earths. The large, spherical *s* orbitals of main-group cations exhibit strong and isotropic overlaps, leading to dispersive conduction bands and correspondingly light electron masses (Fig. 1c). In contrast, the valence band is largely composed of localized anion 2p orbitals, resulting in generally heavier effective masses for holes than for electrons. However, hole masses decrease with increasing anion packing density, which increases the inter-anion 2p orbital overlap (Fig, 1d). Exceptions arise when cation *s* or *d* orbitals contribute to the valence bands, as seen in rs-ZnO, wz-ZnO, and GaN, for which mixed orbital character facilitates hole delocalization.

Moreover, the crystal structure plays an important role in shaping the band dispersions. Highly symmetric and densely packed crystal structures with simple stoichiometries (e.g., 1:1 or 1:2), such as zinc blende (zb), rocksalt (rs), wurtzite (wz), and rutile (r), promote strong orbital overlap and reduce band folding at the Brillouin-zone edges, resulting in lighter hole masses (Fig. 1b). Among these, octahedrally coordinated structures (rocksalt, rutile) yield higher anion packing densities than tetrahedrally coordinated ones (wurtzite, zincblende), further enhancing hole transport.

Overall, both electron and hole masses tend to increase with increasing band gap (Figs. 1e-f). Nonetheless, several oxides of light main-group cations (such as BeO polytypes, MgO, $\gamma$-Al$_2$O$_3$ and r-SiO$_2$) defy this trend due to their dense packing, exhibiting ultra-wide band gaps above 7 eV combined with relatively light electron masses (< 0.7 $m_e$) and hole masses (< 2 $m_e$). In particular, BeO shows multiple promising polytypes: while the ground-state structure is wz, the zb phase lies only ~0.01 eV/atom above the convex hull and is therefore also stable. And though the rs-BeO is unstable at ambient conditions, it can be synthesized from the wurtzite phase under high pressure using a diamond anvil cell[13,14] Alternatively, it can be stabilized at ambient pressure by forming solid solutions with rs-MgO, which shares the same crystal structure and similar cation size, enabling octahedral BeO within a stable host lattice.

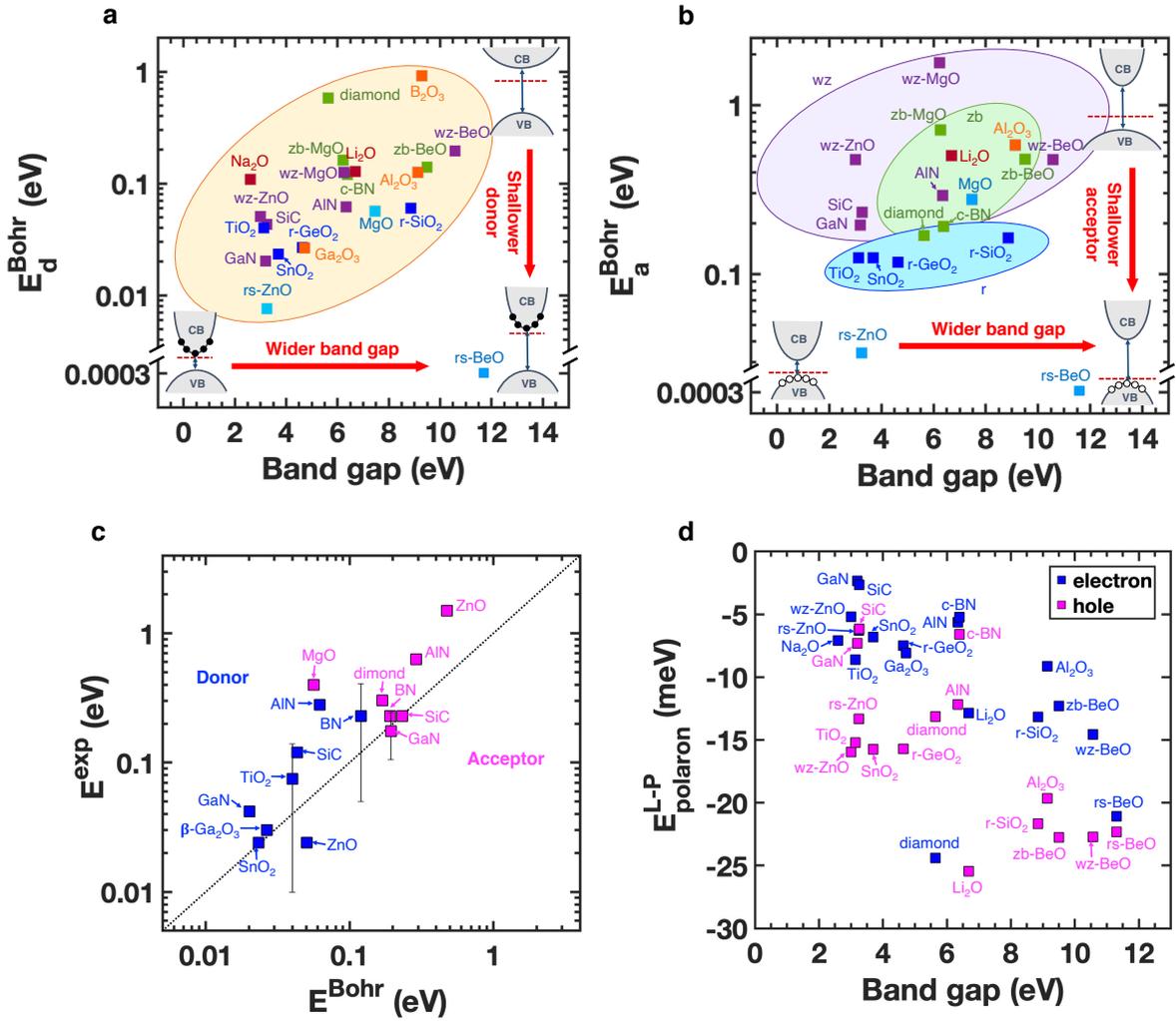

**Fig. 2. Dopant ionization energy and polaron dissociation energy versus band gap. a-b,** The Bohr-model donor ($E_d^{Bohr}$) and acceptor ($E_a^{Bohr}$) ionization energies identify several unexplored ultra-wide-gap materials with shallow dopants such as MgO, r-SiO$_2$, and BeO polytypes. **c,** The model in general yields accurate predictions of experimental shallow donor[10,15–23] and acceptor[10,15–17,24,25] ionization energies. **d,** The polaron dissociation energy for both electrons and holes ($E_{polaron}^{L-P}$) is larger for materials with wider band gaps.

We next identify UWBG materials that can host shallow dopants and exhibit low polaron binding energy. By combining the hydrogenic Bohr model with our calculated effective masses (Fig. 1e-f) and dielectric constants (Extended Data Fig. 2), we evaluate dopant ionization energies across a range of band gaps and find good agreement with experimental data for known semiconductors (Fig. 2a-c). As expected, donors generally become deeper with increasing gap (Fig. 2a), yet our model accurately recovers established n-type semiconductors such as GaN, SiC, AlN, wz-ZnO, $TiO_2$, $SnO_2$ and $\beta$-$Ga_2O_3$. In addition, we identify several less explored materials, including rs-ZnO, r-$GeO_2$, rs-MgO, r-$SiO_2$, and the BeO polytypes, with shallow donor ionization energies (< 0.21 eV). Notably, Ga-doped rs-ZnO is known to exhibit high conductivity under high pressure.[26] Moreover, r-$GeO_2$ has been predicted to host shallow donors,[27] a prediction that has since been confirmed by experiment,[28] which further affirms the validity of our model.

The model predictions for acceptor ionization energies (Fig. 2b) show a similar increasing trend with increasing gap but are generally higher than for donors due to the heavier hole effective masses. However, rutile-structured binary oxides emerge as a notable exception: their model ionization energies remain nearly constant despite increasing band gap. This unusual trend stems from a consistent reduction in lattice constant and a corresponding increase in oxygen packing density in wider-gap compounds, which results in lighter hole effective masses. Remarkably, r-$SiO_2$ combines an extreme band gap of 8.85 eV with a lower predicted acceptor ionization energy than both diamond and c-BN,[29] which are known p-type UWBG materials. Rs-BeO also emerges as a candidate extreme-gap material with shallow dopants of both types, due to its unusually high dielectric constant ($\varepsilon_0$ = 167). However, our phonon calculations revealed imaginary frequencies for rs-BeO, indicating dynamic instability under ambient conditions. Therefore, we consider rs-BeO only as the end member of the rs-$Mg_xBe_{1-x}O$ alloy series, which we demonstrate to be stable for $x$ greater than 0.83 based on our calculation of the enthalpy of mixing in Extended Data Fig. 3.

Finally, the polaron binding energies predicted using the Landau-Pekar model[8] generally increase with band gap, but remain below the thermal energy ($k_B T$) at room temperature (Fig. 2d), suggesting that charge carriers are unlikely to self-trap. In the following sections, we validate these model predictions using atomistic defect and polaron calculations from first principles.

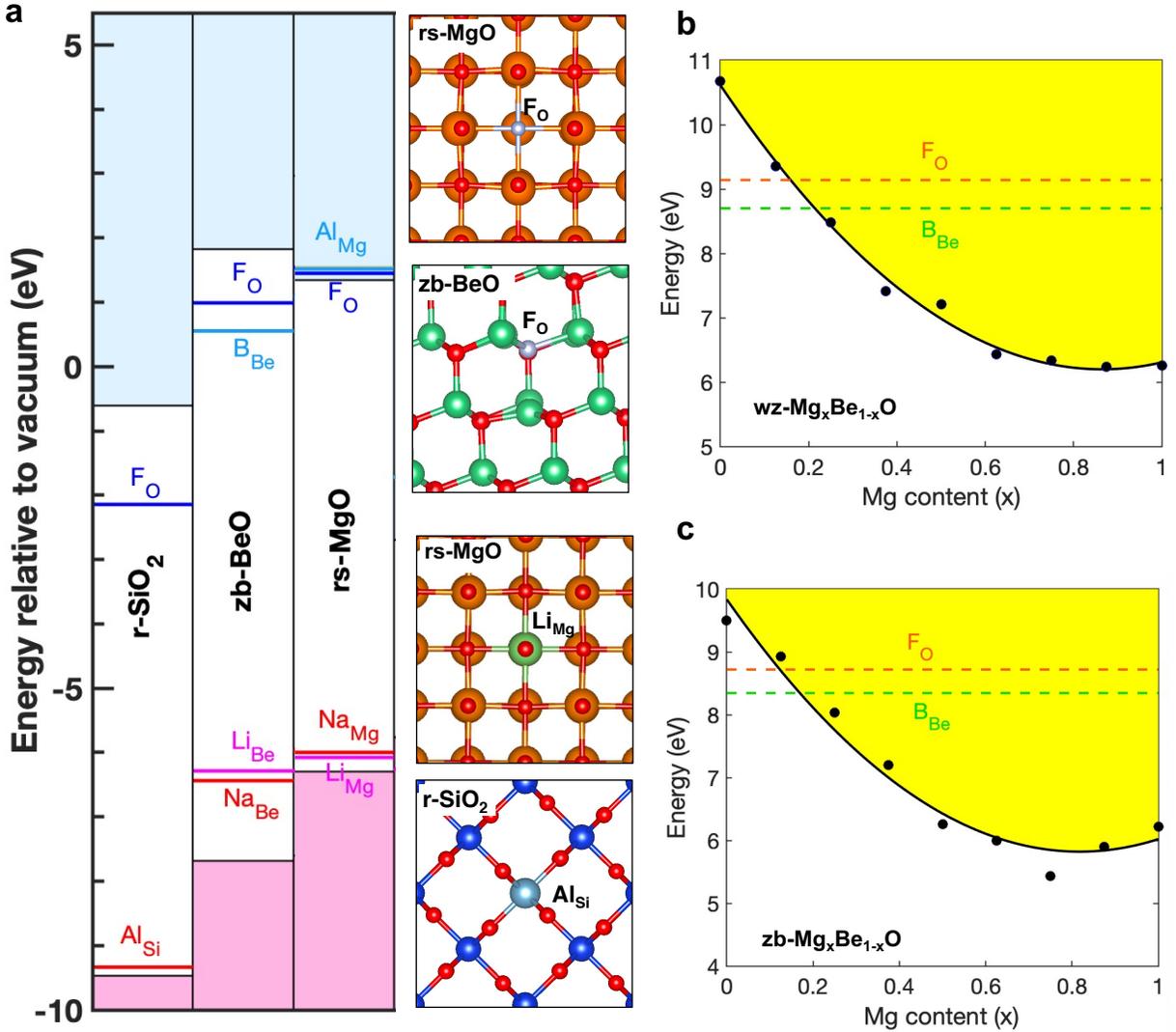

**Fi**gure ... **dop**a... **l**uctors. **a,** (left) Absolute band positions relative to vacuum and dopant energy levels of r-SiO$_2$, zb-BeO and rs-MgO. (right) The defect configuration after relaxation for a donor impurity in rs-MgO and zb-BeO and for an acceptor impurity in rs-MgO and r-SiO$_2$. Despite its ultra-wide band gap (8.87 eV, 9.50 eV and 7.64 eV), r-SiO$_2$, zb-BeO and MgO host shallow dopants without formation of DX- or AX-centers. **b-c,** The level of F$_O$ and B$_{Be}$ obtained from calculations in **b**, wz-BeO and **c**, zb-BeO compared with the CBM energy of wz- and zb-(Mg$_x$Be$_{1-x}$)O alloys, respectively. The CBM energy (shaded yellow) is plotted with respect to the VBM energy of wz- and zb-BeO and obtained by fitting the raw data (black dot) as a function of Mg content.

We next performed atomistic defect calculation using hybrid density functional theory to assess dopability of the UWBG oxides predicted by the Bohr model, and to identify dopant impurities and their atomic configuration. While substitutional impurity atoms may act as shallow dopants in an undistorted lattice, their incorporation can induce strong lattice distortions – known as polaronic effects – that localize the additional carriers and raise the ionization energy. For instance, acceptor dopants in wide-band-gap oxides frequently lead to localized lattice distortions that inhibit p-type doping by trapping holes as polarons.[27,30,31] Moreover, these distortions may break local bonds and generate midgap dangling-bond states capable of capturing an additional carrier, thereby converting the dopants into compensating defects – such as the deep DX-centers that hinder n-type doping in AlGaN.[32] To obtain accurate ionization energies, we explicitly modeled distorted atomic configurations for each dopant (followed by structural relaxation) and compared with the undistorted substitution to assess their relative stability. We find that the tendency for polaron formation depends on both the host material and the impurity species. Donor dopants tend to form DX centers in the BeO polytypes, r-SiO$_2$, and Al$_2$O$_3$, with the notable exception of F$_O$ in zb-BeO (Extended Data Fig. 4c and 5). This exception arises from the absence of third-nearest-neighbor interactions in the zincblende phase, a consequence of its different stacking sequence compared to wurtzite (Extended Data Fig. 4e and 5).[33] In contrast, for acceptor dopants, AX centers are not stable in MgO, r-SiO$_2$, and the BeO polytypes. In these cases, the 0/–1 transition occurs with relatively low ionization energy (Fig. 3 and Extended Data Fig. 4b, d, f, h).

We next calculated the donor and acceptor ionization energies for those impurities that do not promote the formation of DX centers. In rs-MgO, both for F$_O$ and Al$_{Mg}$ are predicted to be effective shallow donors (<0.07 eV), while Li$_{Mg}$ (0.21 eV) or Na$_{Mg}$ (0.29 eV) act as shallow acceptors (Fig 3a and Extended Data Fig. 4b). Although the n-type doping of rs-MgO has not been demonstrated experimentally – likely due to compensation by Mg vacancies and H interstitials – p-type conduction has been reported in both Li- and Na-doped rs-MgO.[23,34] This observation supports our predictions, despite weak compensation by oxygen vacancies and Li interstitials. We further identify r-SiO$_2$ as an ultra-wide-band-gap (8.87 eV) material capable of supporting shallow Al$_{Si}$ acceptors with an ionization energy of only 0.13 eV (Fig. 3a and Extended Data Fig. 4h). This finding aligns with prior work by Lyons et al., who showed that r-SiO$_2$ can accommodate shallow acceptors such as Al and Ga, and that self-trapped holes are energetically unfavorable.[29] F$_O$ is predicted to act as a deep donor in zb-BeO, with an ionization energy of 0.78 eV (Extended Data Fig. 4c). However, it does not form DX centers, enabling degenerate doping at a feasible concentration of approximately 6.4×10$^{19}$ cm$^{-3}$, owing to the light electron effective mass of zb-BeO.

We further propose that alloying zb-BeO and wz-BeO with MgO can lower the energy of the conduction band edge, thereby making the ionization energy of F$_O$ shallower and enabling n-type doping. From the band calculation of wz-Mg$_x$Be$_{1-x}$O, at 16% doping of Mg (corresponding to a band gap of 9.14 eV), the transition level of F$_O$ crosses the conduction band edge (Fig. 3b), indicating the potential for fully ionized donors. In the case of zb-Mg$_x$Be$_{1-x}$O, at 13% alloying with

Mg (corresponding to a band gap of 8.71 eV), the +1/0 transition level of $F_O$ crosses the conduction band edge (Fig. 3c), In contrast, acceptor dopants in zb-BeO and wz-BeO are predicted to be deep, with ionization energies ranging from 1.24 eV to 1.65 eV (Extended Data Fig. 4d, f), precluding p-type doping.

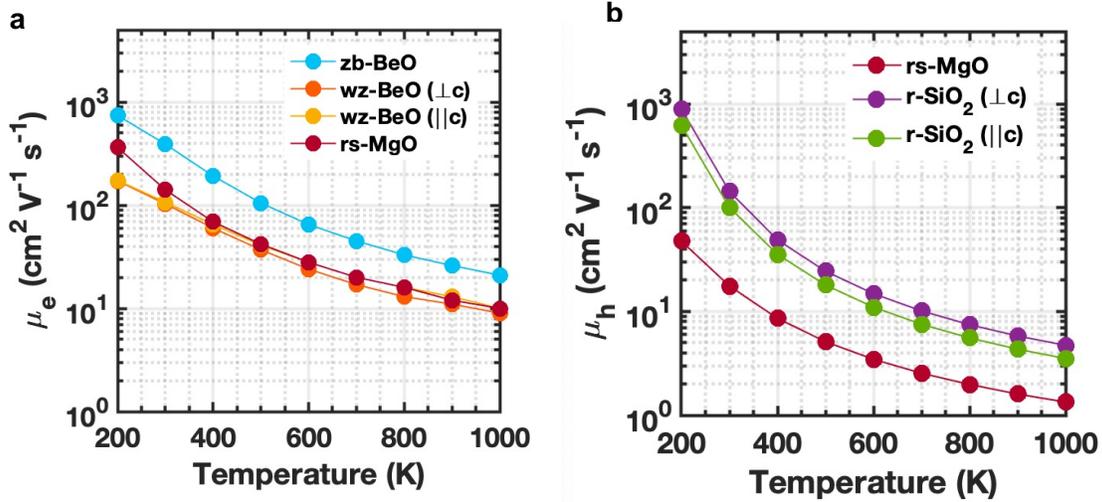

**Fig. 4. Mobility of free carriers in extreme-band-gap semiconductors from first-principles calculations. a,** Electron mobility of zb-BeO, wz-BeO, and rs-MgO, and **b,** hole mobility of rs-MgO and r-SiO$_2$ as a function of temperature.

Last, we apply atomistic calculations to evaluate polarons binding energies and carrier mobilities in the predicted dopable semiconductors from first principles. To study the possibility of charge carrier self-trapping in our candidate materials, we apply the *ab* initio theory of polarons developed in Refs.[8,35] (see Methods). Our calculations predict the formation of large Fröhlich-type polarons with formation energies of 37 meV, 22 meV, and 84 meV in n-type zb-BeO, n-type rs-MgO, and p-type r-SiO$_2$, respectively (see Extended Data Fig.6a-l). The low formation energies indicate that polarons effectively behave as free carriers in these materials. Our results also show that holes tend to localize into stable small polarons with formation energies of 370 meV in p-type rs-MgO, in agreement with previous theoretical work.[36–39] However, we calculate relatively low hopping barriers of 90 meV, which allow for temperature-activated polaronic transport with mobilities of ~0.25 cm$^2$/V·s at room temperature (see Extended Data Fig. 6m-p). In addition, the calculated phonon-limited electron and hole mobilities are shown in Fig. 4a-b. At 300 K, the calculated electron mobility is 141 cm$^2$/V·s for MgO, which is comparable to the mobilities of $\beta$-Ga$_2$O$_3$ or high Al content Al$_x$Ga$_{1-x}$N (x = 0.4 – 0.8),[40,41] and 391 cm$^2$/V·s for zb-BeO. On the other hand, the calculated room-temperature mobility of free holes (i.e., not bound as polarons) is 17 cm$^2$/V·s for MgO. The calculated hole mobility of r-SiO$_2$ is 100 cm$^2$/V·s along the *a* direction and 143 cm$^2$/V·s along the *c* direction. Combined with the prediction for shallow dopants and large polarons, we determine that zb-BeO and rs-MgO can host shallow donors and r-SiO$_2$ shallow acceptors with reasonably high carrier mobility. The key properties of these extreme-band-gap binary semiconductors, along with the dopable zb- and wz- Mg$_x$Be$_{1-x}$O alloys, are summarized in Table I.

**Table I. Key material parameters of extreme-band-gap semiconductors identified in this work.** The electron and hole mobilities for all materials are reported at room temperature.

| Material | $E_g$ (eV) | Doping | $\varepsilon_0$ | $\mu_{e/h}$ (cm$^2$/V·s) | $E_{d/a}$ (eV) | $E_{polaron}$ (eV) |
|---|---|---|---|---|---|---|
| rs-MgO | 7.47 | n-type | 9.28 | 141 | <0.07 | 0.022 |
|  |  | p-type |  | 17 | 0.21 | 0.370 |
| r-SiO$_2$ | 8.85 | p-type | 10.00 | 100 ($\perp$c), 143 ($\parallel$c) | 0.13 | 0.084 |
| zb-BeO | 9.50 | n-type | 6.67 | 391 | 0.78* | 0.037 |
| zb-Mg$_x$Be$_{1-x}$O, x > 13% | < 8.71 | n-type |  |  | 0.140-0.162 |  |
| wz-Mg$_x$Be$_{1-x}$O, x > 19% | < 9.1 | n-type |  |  | 0.126-0.195 |  |

* zb-BeO can be degenerately doped with F$_O$ at a density higher than 6.4×10$^{19}$ cm$^{-3}$

Previous efforts[42–46] to assess semiconductor dopability have primarily focused on the alignment of the conduction and valence band energies with respect to vacuum in order to determine dopant compensation by native defects such as vacancies and interstitials under thermodynamic equilibrium. However, a variety of nonequilibrium experimental techniques have been developed recently to control the Fermi level during growth or annealing, with demonstrated success in suppressing compensation. These methods include above-gap illumination,[47] junction-assisted epitaxy,[48] and co-doping with mobile compensators that can subsequently be removed by annealing[49]. They operate by preventing the shift of the Fermi level closer to the band edges upon doping, and can thus suppress the effects of unfavorable band alignment on compensation. Consequently, we argue that the fundamental limit to dopability in UWBG semiconductors is not compensation, but rather the ability to ionize dopants effectively. This perspective redefines the design rules for UWBG semiconductors by focusing on intrinsic material properties—such as effective mass and dielectric screening—as the key descriptors and enablers of dopability and efficient carrier transport.

The discovery of extreme-gap semiconductors – materials with band gaps exceeding that of AlN, currently the widest-gap semiconductor – could pave the way for groundbreaking technologies in power electronics, deep-UV optoelectronics, and related technologies. Materials with such wide band gaps can directly emit vacuum ultraviolet (VUV) radiation (< 200 nm), which is currently produced using large, gas-based excimer lasers. VUV light also plays a critical role in high-resolution photolithography for nanoscale fabrication and in fluorescence spectroscopy for molecular analysis. In addition, extreme-gap semiconductors offer significantly higher dielectric breakdown fields, enabling compact and energy-efficient power-electronic devices and supporting wider bandwidths for high-frequency applications.

In summary, we have identified the structural and chemical design principles that enable materials to combine ultra-wide band gaps with semiconducting behavior. Our discovery of shallow dopants and mobile carriers in materials such as rs-MgO, r-SiO$_2$, and zb-BeO – with band gaps as wide as 9.50 eV, i.e. extreme-band-gap semiconductors that extend into the band gap range characteristic of insulators, addresses a longstanding question in materials science about the maximum attainable band gap for semiconductors. We find that the effective mass is the key descriptor that governs dopant ionization, carrier mobility, and polaron formation, and thus ultimately defines the boundaries of semiconducting behavior. These findings demonstrate that there is no inherent upper bound to the band gap of semiconductors, as long as dopants can be ionized and charge carriers remain mobile. Our predictive framework for dopant ionization, validated by experiment evidence across a range of UWBG materials, provides a powerful platform for systematically discovering and optimizing novel semiconductors that meet the requirements of future transparent conductors, power electronics, and UV optoelectronics applications.

**Methods**

**Band structure and band alignment.** We performed first-principles calculations based on hybrid density function theory using the Vienna Ab initio Simulation Package (VASP)[50,51]. We first calculate the crystal structure and electronic band structure of the oxides, nitrides, and carbides listed in Figure 1 using the Heyd-Scuseria-Ernzerhof (HSE06) functional[52] and the projected augmented wave (PAW) method. All structures are fully relaxed using the quasi-Newton algorithm with a maximum force criterion of 0.01 eV/Å and fully converged cutoff energy and Brillouin zone sampling grids are used for each material. The amount of Hartree-Fock exchange was adjusted separately for each material between 25–35% to bring its calculated band gap close to the experimental value. The mixing ratio and the resulting structural parameters and band gaps of each material are listed in Extended Data Table I. Though the band gap of rutile $SiO_2$ has not been experimentally measured, we used 35 % Hartree-Fock exchange for rutile $SiO_2$ since the calculated the band gap of quartz $SiO_2$ using 35% Hartree-Fock exchange (8.89 eV) is close to the experimental value (8.8 eV)[53]. Static dielectric constants for all materials are calculated with density functional perturbation theory[54] using the Perdew-Burke-Ernzerhof (PBE) exchange-correlation functional, and listed for each material in Extended Data Table II along with the carrier effective masses and the atomic orbital characters of the valence/conduction band extrema. The calculated HSE06-band structures for all materials studied in this paper are shown in Extended Data Fig. 1 **a-u**.

To align the calculated band structures for MgO, r-$SiO_2$, and the BeO polytypes to vacuum, we generated slab structures using the bulk relaxed lattice constants. Non-polar planes of the highest atomic density (lowest surface energy) are exposed to vacuum: (001) for MgO and rs-BeO, (110) for r-$SiO_2$ and zb-BeO, and (100) for wz-BeO. All slabs consist of 12 layers of atomic planes and a 12 Å-thick vacuum region along the out-of-plane direction. We then calculated the plane-averaged electrostatic potential without relaxing the ions, and use the difference between the average electrostatic potential inside the slab and the vacuum region to align the bulk bands of each material to the vacuum level.[55]

**Effective masses**. Based on our calculated band structure using HSE06 functional, we obtained the hole and electron effective mass by fitting the valence and conduction bands near the extrema with the hyperbolic equation, $E(k) = \frac{\mp 1 \pm \sqrt{1+(4\alpha \hbar k)/(2m^*)}}{2\alpha} + E_1$, where $E(k)$ is band energy as a function of crystal momentum $k$, $\alpha$ is the non-parabolicity fitting parameter, $\hbar$ is the reduced Planck constant, $m^*$ is the electron (–/+) or hole (+/–) effective mass, and $E_1$ is the VBM (–/+) or CBM (+/–) value. For each material, we obtained effective masses along the longitudinal and transverse k-paths at the band edge and the high symmetry k-path in the Brillouin zone is determined by the ref.[56] The spacing for each k-path used for HSE06 band calculation is 50 and 10 - 20 data points near the band extrema were used for fitting the hyperbolic equation. The effective masses are then averaged by the geometric mean (i.e., effective mass for density of state calculations). Effective masses from degenerate bands are averaged by arithmetic mean.

**Defects and dopants**. We calculated the formation energy of a point defect $D$ in charge state $q$, which is defined by[57]

$$E^f(D^q) = E_{tot}(D^q) - E_{tot}(bulk) - \sum n_i(E_i + \mu_i) + q(E_F + E_v) + E_{corr}(D^q),$$

where $E_{tot}(D^q)$ is the total energy of a supercell with a point defect, $E_{tot}(bulk)$ is the total energy of a reference supercell without a point defect, $n_i$ is the number of defect atoms added to or removed from the supercell, $E_i$ is the energy per atom in its elemental phase, $\mu_i$ is the chemical potential, $E_F$ is the Fermi level, $E_v$ is the valence-band maximum, and $E_{corr}(D^q)$ is the correction energy arising from the finite supercell size, which is computed with the SXDFECTALIGN code.[58] We modeled a 72-atom supercell for r-SiO$_2$, and wz-BeO and a 64-atom supercell for rs-BeO, zb-BeO and MgO and used 2 × 2 × 2 Γ-centered Brillouin-zone sampling mesh for all supercells. In the case for intentional dopants, to predict more accurate value of ionization energies, we used 216-atom supercell for rs-BeO, zb-BeO and MgO and 192-atom supercell for wz-BeO and used 2

× 2 × 2 Γ-centered Brillouin-zone for all these materials. We considered two limits of growth conditions, which are the extreme O-rich/cation-poor and cation-rich/O-poor conditions. We also calculated the formation energy of secondary phases to limit the chemical potentials of the impurities. All defect calculations were performed using HSE06 functional.

**Polarons.** Polaron calculations have been performed applying the methodology described in Refs.[59,60] as implemented in the EPW code.[61] Density functional theory calculations are performed with Quantum ESPRESSO[62] within Perdew-Burke-Ernzerhof generalized gradient approximation,[63] using optimized norm-conserving Vanderbilt pseudopotentials[64,65] and plane waves with a kinetic energy cutoff of 80 Ry. Phonon frequencies and electron-phonon matrix elements are computed within density functional perturbation theory[66] on coarse momentum grids of $10 \times 10 \times 10$, $12 \times 12 \times 12$ and $6 \times 6 \times 6$ k and q-points for zb-BeO, rs-MgO, and r-SiO$_2$, respectively, and are interpolated to dense grids by means of Wannier-Fourier interpolation.[67,68] The method presented in Ref.[69] is used to deal with the long-range part of the electron-phonon matrix elements, and quadrupole corrections are also included for zb-BeO and r-SiO$_2$. The quadrupole tensors have been computed using density functional perturbation theory[70] as implemented in the ABINIT software.[71] A more detailed analysis of the properties of this polaron is reported elsewhere.[72]

**Mobility.** We performed quasiparticle mobility calculations for rs-MgO, zb-BeO, wz-BeO, and rutile SiO$_2$ starting from the local density approximation exchange-correlation functional[73] within Quantum ESPRESSO.[74] Phonon frequencies were calculated using density functional perturbation theory[66] on a 6×6×6 Brillouin-zone (BZ) grid for rs-MgO and zb-BeO, an 8×8×6 BZ grid for wz-BeO, and an 8×8×12 BZ grid for rutile SiO$_2$. Quasiparticle energies were calculated with the G$_0$W$_0$ method[75] as implemented in BerkeleyGW[76] using a 6×6×6 BZ sampling grid for rs-MgO and zb-

BeO, an 8×8×6 BZ grid for wz-BeO, and an 8×8×12 BZ grid for rutile SiO$_2$, a screening plane-wave cutoff energy of 40 Ry, and a summation over unoccupied states up to 20 Ry within the static remainder approach.[77] Electron-phonon coupling matrix elements were evaluated within the Electron-Phonon-Wannier (EPW) code[78] and interpolated to fine electron and phonon BZ sampling meshes up to 120×120×120 for rs-MgO and zb-BeO, 96×96×72 for wz-BeO, and 48×48×72 for rutile SiO$_2$. The phonon-limited electron mobility was evaluated as a function of temperature with the iterative Boltzmann Transport Equation method[79,80] for states within a 0.5 eV energy window above the conduction band minimum. The imaginary self energy of electrons due to the electron-phonon interaction (i.e., the inverse of the carrier lifetime), Im(Σ), was computed for states within a 0.25 eV energy window above the conduction band minimum.

**Methods References**

**Data availability**

The data that support the findings of this study are available upon reasonable request.

**Acknowledgements**

We acknowledge fruitful discussions with C. G. Van de Walle and N. A. Modine. This work was supported by the Computational Materials Sciences Program funded by the U.S. Department of Energy, Office of Science, Basic Energy Sciences, under Award DE-SC0020129. This research used resources of the National Energy Research Scientific Computing Center, a DOE Office of Science User Facility supported by the Office of Science of the U.S. Department of Energy under Contract No. DE-AC02-05CH11231. We acknowledge the Texas Advanced Computing Center at The University of Texas at Austin for providing access to Frontera and Lonestar6 (http://www.tacc.utexas.edu).

**Author contributions**

S.C., J.T.H. and E.K. conceptualized the study. S.C. and E.K. performed and analyzed the band-structure and defect calculations. S.C. and Y.-C. H. performed and analyzed the thermodynamic and band-alignment calculations for MgBeO alloys. J. L.-B., K. L, and F.G. performed and analyzed the polaron calculations. N.S., K.A.M., A.W., X.Z., and E.K. performed and analyzed the mobility calculations. S.C., J.T.H., and E.K. prepared the manuscript, which was then revised by all co-authors. All authors have given approval for the final version of the manuscript.

**Competing interest declaration** The authors declare no competing interests.

**Additional information**

**Correspondence and requests for materials** should be addressed to E. Kioupakis and S. Chae.

**Reprints and permissions information** is available at www.nature.com/reprints

**Extended Data Table I. The information of crystal structure and electronic structure.** $E_{g,exp}$ and $E_{g,HSE}$ refer to experimental and HSE06-calculated band gap.

| Material composition | Crystal structure | Lattice parameter (Å) | | | Packing density (atoms/nm$^3$) | | $E_{g,exp}$ (eV) | $E_{g,HSE}$ (eV) | HF mixing (%) |
|---|---|---|---|---|---|---|---|---|---|
| | | *a* | b | c | Cation | Anion | | | |
| GaN | Wurtzite | 3.199 | 3.199 | 5.198 | 43.4 | 43.4 | 3.28 | 3.194 | 25 |
| AlN | Wurtzite | 3.091 | 3.091 | 4.945 | 48.8 | 48.8 | 6.2 | 6.340 | 35 |
| SiC | Wurtzite | 3.067 | 3.067 | 10.036 | 48.9 | 48.9 | 3.2 | 3.255 | 27 |
| BeO | Wurtzite | 2.675 | 2.675 | 4.340 | 73.5 | 73.5 | | 10.570 | 35 |
| ZnO | Wurtzite | 3.224 | 3.224 | 5.177 | 40.2 | 40.2 | 3.3 | 3.000 | 32 |
| MgO | Wurtzite | 3.261 | 3.261 | 4.917 | 44.2 | 44.2 | | 6.256 | 35 |
| SiO$_2$ | Rutile | 4.157 | 4.157 | 2.656 | 43.6 | 87.2 | | 8.851 | 35 |
| GeO$_2$ | Rutile | 4.395 | 4.395 | 2.866 | 36.1 | 72.2 | 4.68 | 4.640 | 35 |
| SnO$_2$ | Rutile | 4.711 | 4.711 | 3.169 | 28.0 | 56.0 | 3.6 | 3.687 | 32 |
| TiO$_2$ | Rutile | 4.597 | 4.597 | 2.946 | 32.1 | 64.2 | 3.05 | 3.131 | 30 |
| BN | Zincblende | 3.586 | 3.586 | 3.586 | 86.8 | 86.4 | 6.45 | 6.387 | 35 |
| Diamond | Zincblende | 3.547 | 3.547 | 3.547 | 90.0 | 90.0 | 5.6 | 5.642 | 30 |
| BeO | Zincblende | 3.757 | 3.757 | 3.757 | 75.4 | 75.4 | | 9.495 | 35 |
| MgO | Zincblende | 4.503 | 4.503 | 4.503 | 43.8 | 43.8 | | 6.219 | 35 |
| MgO | Rocksalt | 4.177 | 4.177 | 4.177 | 54.9 | 54.9 | 7.77 | 7.468 | 35 |
| BeO | Rocksalt | 3.589 | 3.589 | 3.589 | 85.4 | 85.4 | | 11.620 | 35 |
| ZnO | Rocksalt | 4.230 | 4.230 | 4.230 | 49.0 | 49.0 | | 3.245 | 30 |
| Li$_2$O | Fluorite | 4.572 | 4.572 | 4.572 | 83.6 | 41.8 | 7-7.5 | 6.687 | 25 |
| Na$_2$O | Fluorite | 5.457 | 5.457 | 5.457 | 49.2 | 24.6 | | 2.598 | 25 |
| B$_2$O$_3$ | Trigonal | 4.332 | 4.332 | 8.411 | 43.9 | 65.9 | | 9.290 | 35 |
| Al$_2$O$_3$ | Corundum | 4.729 | 4.729 | 12.911 | 48.0 | 72.0 | 9.9 | 9.134 | 35 |
| Ga$_2$O$_3$ | Monoclinic | 12.122 | 3.026 | 5.792 | 38.8 | 58.2 | 4.48 | 4.717 | 30 |

**Extended Data Table II. Dielectric constant ($\varepsilon_0$), electron and hole effective masses ($m_e^*$ and $m_h^*$), and the projection of orbitals constituting VBM and CBM.**

| Material composition | Crystal structure | $\varepsilon_0$ | $\varepsilon_\infty$ | $m_e^*$ ($m_e$) | $m_h^*$ ($m_e$) | VBM orbital | | | CBM orbital | | |
|---|---|---|---|---|---|---|---|---|---|---|---|
| | | | | | | s | p | d | s | p | d |
| GaN | Wurtzite | 9.71 | 5.72 | 0.140 | 1.350 | 0.000 | 0.682 | 0.166 | 0.628 | 0.002 | 0.000 |
| AlN | Wurtzite | 8.48 | 4.59 | 0.328 | 1.539 | 0.000 | 0.624 | 0.000 | 0.484 | 0.000 | 0.000 |
| SiC | Wurtzite | 10.55 | 7.20 | 0.353 | 1.901 | 0.000 | 0.530 | 0.000 | 0.458 | 0.026 | 0.000 |
| BeO | Wurtzite | 6.61 | 3.11 | 0.628 | 1.526 | 0.000 | 0.834 | 0.000 | 0.470 | 0.001 | 0.000 |
| ZnO | Wurtzite | 8.62 | 4.05 | 0.276 | 2.598 | 0.000 | 0.635 | 0.222 | 0.363 | 0.177 | 0.005 |
| MgO | Wurtzite | 6.276 | 3.02 | 0.365 | 2.056 | 0.000 | 0.732 | 0.024 | 0.323 | 0.000 | 0.000 |
| SiO$_2$ | Rutile | 10.00 | 3.46 | 0.444 | 1.199 | 0.000 | 0.896 | 0.002 | 0.842 | 0.006 | 0.008 |
| GeO$_2$ | Rutile | 11.92 | 4.25 | 0.280 | 1.230 | 0.000 | 0.876 | 0.001 | 0.906 | 0.020 | 0.017 |
| SnO$_2$ | Rutile | 11.96 | 4.19 | 0.245 | 1.310 | 0.000 | 0.842 | 0.001 | 0.870 | 0.014 | 0.016 |
| TiO$_2$ | Rutile | 17.04 | 6.27 | 0.853 | 2.658 | 0.000 | 0.764 | 0.037 | 0.000 | 0.008 | 0.953 |
| BN | Zincblende | 6.70 | 4.54 | 0.398 | 0.634 | 0.000 | 0.796 | 0.000 | 0.000 | 0.682 | 0.162 |
| Diamond | Zincblende | 5.87 | 5.87 | 1.466 | 0.427 | 0.000 | 0.753 | 0.000 | 0.000 | 0.582 | 0.390 |
| BeO | Zincblende | 6.67 | 3.14 | 0.459 | 1.568 | 0.000 | 0.836 | 0.000 | 0.065 | 0.278 | 0.000 |
| MgO | Zincblende | 5.530 | 3.04 | 0.364 | 4.740 | 0.000 | 0.714 | 0.032 | 0.323 | 0.000 | 0.000 |
| MgO | Rocksalt | 9.28 | 3.15 | 0.358 | 1.753 | 0.000 | 0.765 | 0.000 | 0.512 | 0.000 | 0.000 |
| BeO | Rocksalt | 166.8 | 3.71 | 0.601 | 0.673 | 0.000 | 0.871 | 0.000 | 0.740 | 0.000 | 0.000 |
| ZnO | Rocksalt | 20.60 | 4.50 | 0.238 | 1.063 | 0.000 | 0.620 | 0.239 | 0.697 | 0.000 | 0.000 |
| Li$_2$O | Fluorite | 7.38 | 2.93 | 0.515 | 2.012 | 0.000 | 0.729 | 0.000 | 0.484 | 0.000 | 0.000 |
| Na$_2$O | Fluorite | 6.78 | 3.11 | 0.358 | >10 | 0.000 | 0.647 | 0.000 | 0.400 | 0.000 | 0.000 |
| B$_2$O$_3$ | Trigonal | 5.44 | 2.95 | 2.005 | >10 | 0.001 | 0.729 | 0.000 | 0.014 | 0.595 | 0.215 |
| Al$_2$O$_3$ | Corundum | 6.46 | 3.27 | 0.387 | 1.784 | 0.010 | 0.791 | 0.000 | 0.617 | 0.002 | 0.000 |
| Ga$_2$O$_3$ | Monoclinic | 11.88 | 3.91 | 0.276 | >10 | 0.001 | 0.773 | 0.059 | 0.694 | 0.015 | 0.016 |

**Extended Data Table III. Ionization energies and polaron formation energies in units of meV.** Ionization energies are predicted by Bohr model ($E_d^{Bohr}/E_a^{Bohr}$) and DFT ($E_d^{DFT}/E_a^{DFT}$) and compared with experimentally determined values ($E_d^{Exp}/E_a^{Exp}$). Polaron formation energies are predicted by Landau-Pekar model ($E_{e,\,polaron}^{L-P}/E_{h,\,polaron}^{L-P}$). The asterisk (*) indicates the data from the present work

| Material composition | Crystal structure | $E_d^{Bohr}$ | $E_a^{Bohr}$ | $E_d^{Exp}$ | $E_a^{Exp}$ | $E_d^{DFT}$ | $E_a^{DFT}$ | $E_{e,\,polaron}^{L-P}$ | $E_{h,\,polaron}^{L-P}$ |
|---|---|---|---|---|---|---|---|---|---|
| GaN | Wurtzite | 20 | 195 | 42 | 140 - 210 | 0[1] | 170[2] | -2.4 | -7.3 |
| AlN | Wurtzite | 62 | 291 | 280 | 630 | 150[1] | 500[3] | -5.6 | -12.2 |
| SiC | Wurtzite | 43 | 232 | 120 | 230 | 0[4] | - | -2.7 | -6.2 |
| BeO | Wurtzite | 195 | 475 | - | - | 1435* | 1319* | -14.6 | -22.7 |
| ZnO | Wurtzite | 51 | 475 | 24 – 51 | 1500 | 0[5] | >600[6] | -5.2 | -16.0 |
| MgO | Wurtzite | 126 | 710 | - | - | - | - | - | - |
| SiO$_2$ | Rutile | 60 | 163 | - | - | 1530* | 1410* | -13.2 | -21.7 |
| GeO$_2$ | Rutile | 27 | 118 | - | - | 0[7] | 450[7] | -7.5 | -15.7 |
| SnO$_2$ | Rutile | 23 | 125 | 24 | 256 | 0[8] | 910[8] | -6.8 | -15.8 |
| TiO$_2$ | Rutile | 715 | 2146 | 10 | - | - | - | -8.6 | -15.2 |
| BN | Zincblende | 120 | 192 | 50 – 410 | 230 | 110[9] | 240[10] | -5.2 | -6.6 |
| Diamond | Zincblende | 579 | 168 | - | 302 | 380[11] | 340[11] | -24.4 | -13.2 |
| BeO | Zincblende | 140 | 479 | - | - | 1237* | 1243* | -12.3 | -22.8 |
| MgO | Zincblende | 162 | 2108 | - | - | - | - | - | - |
| MgO | Rocksalt | 56 | 277 | - | 400 | 0* | 214* | -10.8 | -23.9 |
| BeO | Rocksalt | 0.3 | 0.3 | - | - | 2065* | 69* | 21.1 | -22.3 |
| ZnO | Rocksalt | 8 | 34 | - | - | 0[12] | 0[12] | -6.3 | -13.3 |
| Li$_2$O | Fluorite | 129 | 502 | - | - | - | - | -12.9 | -25.4 |

| | | | | | | | | |
|---|---|---|---|---|---|---|---|---|
| Na$_2$O | Fluorite | 109 | - | - | - | - | - | -7.1 | - |
| B$_2$O$_3$ | Trigonal | 919 | - | - | - | - | - | - | - |
| Al$_2$O$_3$ | Corundum | 126 | 581 | - | - | 763* | 1170* | -9.1 | -19.7 |
| Ga$_2$O$_3$ | Monoclinic | 27 | - | 30 | 1100 | 0[13] | >1300[14] | -8.1 | - |

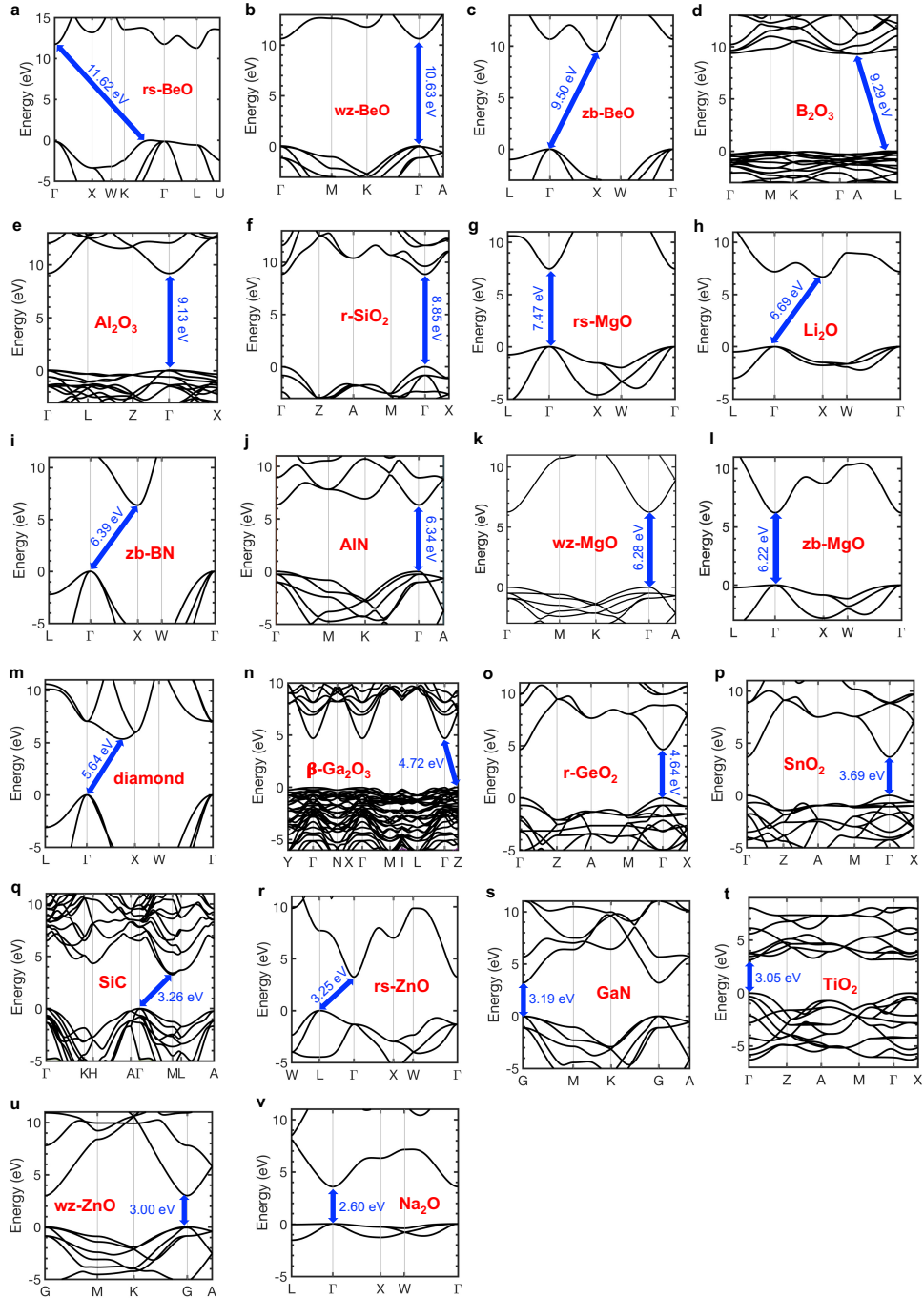

**Extended Data Fig. 1. The HSE06-calculated band structure of binary oxides, nitrides, and carbides composed of light elements in a simple crystal structure.** Arrows indicate the magnitude of the band gap and the location of the valence and conduction band extrema. The materials from **a** to **u** are ordered by decreasing band gap.

**Extended Data Table IV. The calculated effective masses of the materials studied.**

|  | Electron effective mass | Hole effective mass |
|---|---|---|
| GaN | $m_e^*/m_0$ ($\Gamma \to A$) = 0.182<br>$m_e^*/m_0$ ($\Gamma \to M$) = 0.151<br>$m_e^*/m_0$ ($\Gamma \to K$) = 0.100<br>$m_{e,ave}^*/m_0$ = 0.140 | $m_h^*/m_0$ ($\Gamma \to A$) = 1.874<br>$m_h^*/m_0$ ($\Gamma \to M$) = 1.508<br>$m_h^*/m_0$ ($\Gamma \to K$) = 0.870<br>$m_{h,ave}^*/m_0$ = 1.350 |
| AlN | $m_e^*/m_0$ ($\Gamma \to A$) = 0.298<br>$m_e^*/m_0$ ($\Gamma \to M$) = 0.392<br>$m_e^*/m_0$ ($\Gamma \to K$) = 0.303<br>$m_{e,ave}^*/m_0$ = 0.328 | $m_h^*/m_0$ ($\Gamma \to A$) = 0.250<br>$m_h^*/m_0$ ($\Gamma \to M$) = 4.121<br>$m_h^*/m_0$ ($\Gamma \to K$) = 3.539<br>$m_{h,ave}^*/m_0$ = 1.539 |
| wz-BeO | $m_e^*/m_0$ ($\Gamma \to A$) = 0.521<br>$m_e^*/m_0$ ($\Gamma \to M$) = 0.781<br>$m_e^*/m_0$ ($\Gamma \to K$) = 0.609<br>$m_{e,ave}^*/m_0$ = 0.628 | $m_h^*/m_0$ ($\Gamma \to A$) = 0.362<br>$m_h^*/m_0$ ($\Gamma \to M$) = 3.543<br>$m_h^*/m_0$ ($\Gamma \to K$) = 2.776<br>$m_{h,ave}^*/m_0$ = 1.526 |
| wz-ZnO | $m_e^*/m_0$ ($\Gamma \to A$) = 0.252<br>$m_e^*/m_0$ ($\Gamma \to M$) = 0.334<br>$m_e^*/m_0$ ($\Gamma \to K$) = 0.251<br>$m_{e,ave}^*/m_0$ = 0.276 | $m_h^*/m_0$ ($\Gamma \to A$) = 2.628<br>$m_h^*/m_0$ ($\Gamma \to M$) = 2.994<br>$m_h^*/m_0$ ($\Gamma \to K$) = 2.229<br>$m_{h,ave}^*/m_0$ = 2.598 |
| wz-MgO | $m_e^*/m_0$ ($\Gamma \to A$) = 0.338<br>$m_e^*/m_0$ ($\Gamma \to M$) = 0.444<br>$m_e^*/m_0$ ($\Gamma \to K$) = 0.323<br>$m_{e,ave}^*/m_0$ = 0.365 | $m_h^*/m_0$ ($\Gamma \to A$) = 0.396<br>$m_h^*/m_0$ ($\Gamma \to M$) = 5.592<br>$m_h^*/m_0$ ($\Gamma \to K$) = 3.931<br>$m_{h,ave}^*/m_0$ = 2.056 |
| SiC | $m_e^*/m_0$ ($M \to L$) = 0.291<br>$m_e^*/m_0$ ($M \to \Gamma$) = 0.520<br>$m_{e,ave}^*/m_0$ = 0.353 | $m_h^*/m_0$ ($\Gamma \to A$) = 1.473<br>$m_h^*/m_0$ ($\Gamma \to M$) = 2.825<br>$m_h^*/m_0$ ($\Gamma \to K$) = 1.650<br>$m_{h,ave}^*/m_0$ = 1.901 |
| r-SiO$_2$ | $m_e^*/m_0$ ($\Gamma \to Z$) = 0.367<br>$m_e^*/m_0$ ($\Gamma \to X$) = 0.488<br>$m_{e,ave}^*/m_0$ = 0.444 | $m_h^*/m_0$ ($\Gamma \to Z$) = 1.574<br>$m_h^*/m_0$ ($\Gamma \to X$) = 1.047<br>$m_{h,ave}^*/m_0$ = 1.199 |
| r-GeO$_2$ | $m_e^*/m_0$ ($\Gamma \to Z$) = 0.234<br>$m_e^*/m_0$ ($\Gamma \to X$) = 0.307<br>$m_{e,ave}^*/m_0$ = 0.280 | $m_h^*/m_0$ ($\Gamma \to Z$) = 1.565<br>$m_h^*/m_0$ ($\Gamma \to X$) = 1.091<br>$m_{h,ave}^*/m_0$ = 1.230 |
| SnO$_2$ | $m_e^*/m_0$ ($\Gamma \to Z$) = 0.212<br>$m_e^*/m_0$ ($\Gamma \to X$) = 0.264<br>$m_{e,ave}^*/m_0$ = 0.245 | $m_h^*/m_0$ ($\Gamma \to Z$) = 1.577<br>$m_h^*/m_0$ ($\Gamma \to X$) = 1.194<br>$m_{h,ave}^*/m_0$ = 1.310 |
| r-TiO$_2$ | $m_e^*/m_0$ ($\Gamma \to Z$) = 0.537<br>$m_e^*/m_0$ ($\Gamma \to X$) = 1.076<br>$m_{e,ave}^*/m_0$ = 0.853 | $m_h^*/m_0$ ($\Gamma \to Z$) = 3.290<br>$m_h^*/m_0$ ($\Gamma \to X$) = 2.389<br>$m_{h,ave}^*/m_0$ = 2.658 |
| zb-BN | $m_e^*/m_0$ ($X \to \Gamma$) = 0.808<br>$m_e^*/m_0$ ($X \to W$) = 0.279<br>$m_{e,ave}^*/m_0$ = 0.398 | $m_{hh}^*/m_0$ ($\Gamma \to L$) = 1.144<br>$m_{hh}^*/m_0$ ($\Gamma \to X$) = 0.484<br>$m_{hh}^*/m_0$ ($\Gamma \to W$) = 1.558<br>$m_{hh,ave}^*/m_0$ = 0.952<br>$m_{lh}^*/m_0$ ($\Gamma \to L$) = 1.144<br>$m_{lh}^*/m_0$ ($\Gamma \to X$) = 0.484<br>$m_{lh}^*/m_0$ ($\Gamma \to W$) = 0.484<br>$m_{lh,ave}^*/m_0$ = 0.645<br>$m_{soh}^*/m_0$ ($\Gamma \to L$) = 0.220<br>$m_{soh}^*/m_0$ ($\Gamma \to X$) = 0.462<br>$m_{soh}^*/m_0$ ($\Gamma \to W$) = 0.278<br>$m_{soh,ave}^*/m_0$ = 0.304<br>$m_{h,ave}^*/m_0$ = 0.634 |
| diamond | $m_{e,t}^*/m_0$ (CBM → Γ) = 1.645<br>$m_{e,l}^*/m_0$ = 1.688<br>$m_{e,ave}^*/m_0$ = 1.674 | $m_{hh}^*/m_0$ (Γ → L) = 0.656<br>$m_{hh}^*/m_0$ (Γ → X) = 0.472<br>$m_{hh}^*/m_0$ (Γ → W) = 1.075<br>$m_{hh,ave}^*/m_0$ = 0.693<br>$m_{lh}^*/m_0$ (Γ → L) = 0.656<br>$m_{lh}^*/m_0$ (Γ → X) = 0.282<br>$m_{lh}^*/m_0$ (Γ → W) = 0.283<br>$m_{lh,ave}^*/m_0$ = 0.374<br>$m_{soh}^*/m_0$ (Γ → L) = 0.163<br>$m_{soh}^*/m_0$ (Γ → X) = 0.282<br>$m_{soh}^*/m_0$ (Γ → W) = 0.211<br>$m_{soh,ave}^*/m_0$ = 0.213<br>$m_{h,ave}^*/m_0$ = 0.427 |
| zb-BeO | $m_e^*/m_0$ (X → Γ) = 0.674<br>$m_e^*/m_0$ (X → W) = 0.379<br>$m_{e,ave}^*/m_0$ = 0.459 | $m_{hh}^*/m_0$ (Γ → L) = 3.119<br>$m_{hh}^*/m_0$ (Γ → X) = 1.172<br>$m_{hh}^*/m_0$ (Γ → W) = 3.223<br>$m_{hh,ave}^*/m_0$ = 2.275<br>$m_{lh}^*/m_0$ (Γ → L) = 2.808<br>$m_{lh}^*/m_0$ (Γ → X) = 1.170<br>$m_{lh}^*/m_0$ (Γ → W) = 1.171<br>$m_{lh,ave}^*/m_0$ = 1.567<br>$m_{soh}^*/m_0$ (Γ → L) = 2.806<br>$m_{soh}^*/m_0$ (Γ → X) = 0.543<br>$m_{soh}^*/m_0$ (Γ → W) = 0.419<br>$m_{soh,ave}^*/m_0$ = 0.861<br>$m_{h,ave}^*/m_0$ = 1.568 |
| zb-MgO | $m_e^*/m_0$ (Γ → L) = 0.364<br>$m_e^*/m_0$ (Γ → X) = 0.364<br>$m_e^*/m_0$ (Γ → W) = 0.363<br>$m_{e,ave}^*/m_0$ = 0.364 | $m_{hh}^*/m_0$ (Γ → L) = 9.872<br>$m_{hh}^*/m_0$ (Γ → X) = 3.267<br>$m_{hh}^*/m_0$ (Γ → W) = 10.2<br>$m_{hh,ave}^*/m_0$ = 8.499<br>$m_{lh}^*/m_0$ (Γ → L) = 9.868<br>$m_{lh}^*/m_0$ (Γ → X) = 3.208<br>$m_{lh}^*/m_0$ (Γ → W) = 3.197<br>$m_{lh,ave}^*/m_0$ = 5.252<br>$m_{soh}^*/m_0$ (Γ → L) = 0.430<br>$m_{soh}^*/m_0$ (Γ → X) = 0.521<br>$m_{soh}^*/m_0$ (Γ → W) = 0.466<br>$m_{soh,ave}^*/m_0$ = 0.468<br>$m_{h,ave}^*/m_0$ = 4.740 |
| rs-BeO | $m_e^*/m_0$ (Γ → L) = 0.595<br>$m_e^*/m_0$ (Γ → X) = 0.606<br>$m_e^*/m_0$ (Γ → W) = 0.602<br>$m_{e,ave}^*/m_0$ = 0.601 | $m_{h,t}^*/m_0$ (VBM → Γ) = 2.172<br>$m_{h,l}^*/m_0$ = 0.379<br>$m_{h,ave}^*/m_0$ = 0.678 |
| rs-MgO | $m_e^*/m_0$ (Γ → L) = 0.358<br>$m_e^*/m_0$ (Γ → X) = 0.358<br>$m_e^*/m_0$ (Γ → W) = 0.357<br>$m_{e,ave}^*/m_0$ = 0.358 | $m_{hh}^*/m_0$ (Γ → L) = 3.155<br>$m_{hh}^*/m_0$ (Γ → X) = 1.852<br>$m_{hh}^*/m_0$ (Γ → W) = 3.220<br>$m_{hh,ave}^*/m_0$ = 2.660<br>$m_{lh}^*/m_0$ (Γ → L) = 3.157<br>$m_{lh}^*/m_0$ (Γ → X) = 1.867<br>$m_{lh}^*/m_0$ (Γ → W) = 1.872<br>$m_{lh,ave}^*/m_0$ = 2.226<br>$m_{soh}^*/m_0$ (Γ → L) = 0.346 |
| | | $m_{soh}^*/m_0 (\Gamma \to X) = 0.408$<br>$m_{soh}^*/m_0 (\Gamma \to W) = 0.372$<br>$m_{soh,ave}^*/m_0 = 0.374$<br>$m_{h,ave}^*/m_0 = 1.753$ |
| rs-ZnO | $m_e^*/m_0 (\Gamma \to L) = 0.237$<br>$m_e^*/m_0 (\Gamma \to X) = 0.238$<br>$m_e^*/m_0 (\Gamma \to W) = 0.237$<br>$m_{e,ave}^*/m_0 = 0.238$ | $m_{hh}^*/m_0 (L \to \Gamma) = 2.102$<br>$m_{hh}^*/m_0 (L \to W) = 1.258$<br>$m_{hh,ave}^*/m_0 = 1.493$<br>$m_{lh}^*/m_0 (L \to \Gamma) = 2.091$<br>$m_{lh}^*/m_0 (L \to W) = 0.349$<br>$m_{lh,ave}^*/m_0 = 0.634$<br>$m_{h,ave}^*/m_0 = 1.063$ |
| Li$_2$O | $m_e^*/m_0 (X \to \Gamma) = 0.770$<br>$m_e^*/m_0 (X \to W) = 0.421$<br>$m_{e,ave}^*/m_0 = 0.515$ | $m_{hh}^*/m_0 (\Gamma \to L) = 3.882$<br>$m_{hh}^*/m_0 (\Gamma \to X) = 1.676$<br>$m_{hh}^*/m_0 (\Gamma \to W) = 4.680$<br>$m_{hh,ave}^*/m_0 = 3.123$<br>$m_{lh}^*/m_0 (\Gamma \to L) = 3.884$<br>$m_{lh}^*/m_0 (\Gamma \to X) = 1.671$<br>$m_{lh}^*/m_0 (\Gamma \to W) = 1.670$<br>$m_{lh,ave}^*/m_0 = 2.213$<br>$m_{soh}^*/m_0 (\Gamma \to L) = 0.559$<br>$m_{soh}^*/m_0 (\Gamma \to X) = 0.910$<br>$m_{soh}^*/m_0 (\Gamma \to W) = 0.674$<br>$m_{soh,ave}^*/m_0 = 0.700$<br>$m_{h,ave}^*/m_0 = 2.012$ |
| Na$_2$O | $m_e^*/m_0 (\Gamma \to L) = 0.368$<br>$m_e^*/m_0 (\Gamma \to X) = 0.368$<br>$m_e^*/m_0 (\Gamma \to W) = 0.368$<br>$m_{e,ave}^*/m_0 = 0.368$ | Flat band |
| B$_2$O$_3$ | $m_e^*/m_0 (A \to \Gamma) = 1.694$<br>$m_e^*/m_0 (A \to L) = 2.174$<br>$m_{e,ave}^*/m_0 = 2.001$ | Flat band |
| Al$_2$O$_3$ | $m_e^*/m_0 (\Gamma \to L) = 0.378$<br>$m_e^*/m_0 (\Gamma \to Z) = 0.408$<br>$m_e^*/m_0 (\Gamma \to X) = 0.376$<br>$m_{e,ave}^*/m_0 = 0.387$ | $m_h^*/m_0 (\Gamma \to L) = 3.683$<br>$m_h^*/m_0 (\Gamma \to Z) = 0.316$<br>$m_h^*/m_0 (\Gamma \to X) = 4.876$<br>$m_{h,ave}^*/m_0 = 1.784$ |
| $\beta$-Ga$_2$O$_3$ | $m_e^*/m_0 (\Gamma \to Y) = 0.280$<br>$m_e^*/m_0 (\Gamma \to Z) = 0.271$<br>$m_e^*/m_0 (\Gamma \to L) = 0.277$<br>$m_{e,ave}^*/m_0 = 0.276$ | Flat band |

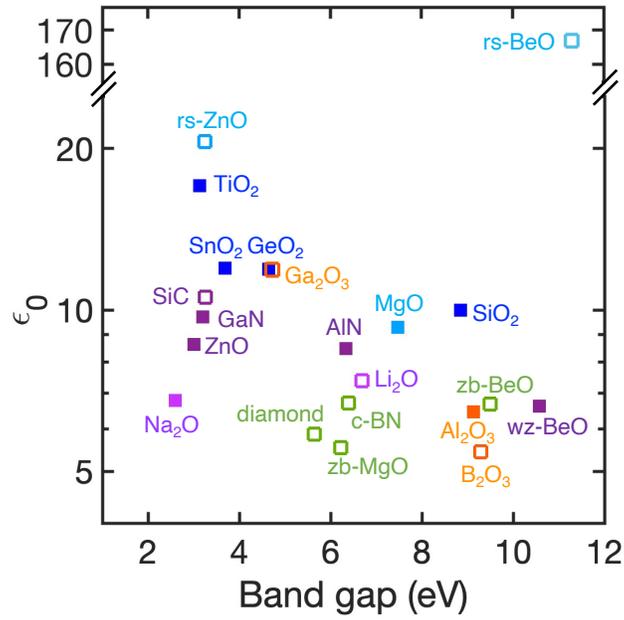

**Extended Data Fig. 2. The HSE06-calculated static dielectric constant of binary oxides, nitrides, and carbides as a function of a band gap.** Dielectric constant generally decreases with increasing band gap.

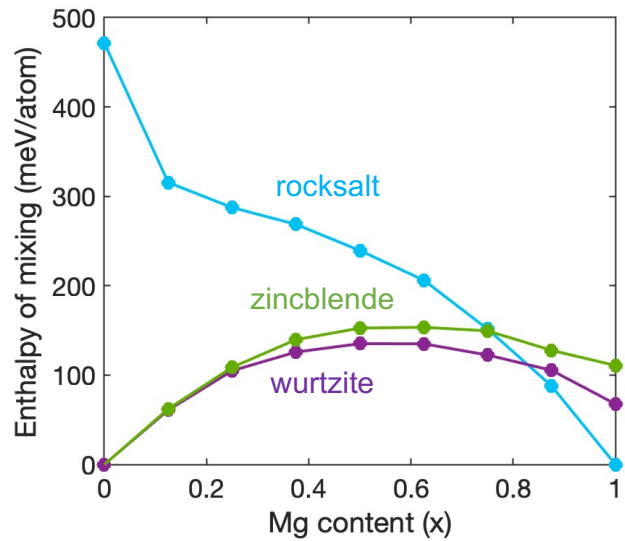

**Extended Data Fig. 3. The enthalpy of mixing of random $(Mg_x,Be_{1-x})O$ alloys** in the rocksalt (skyblue), zincblende (green), and wurtzite (purple) structure.

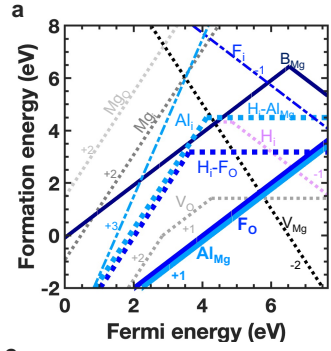
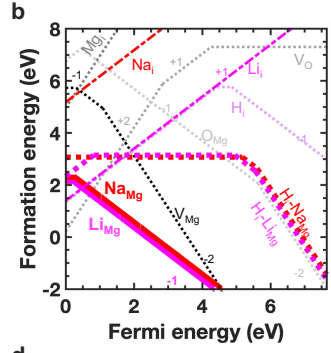
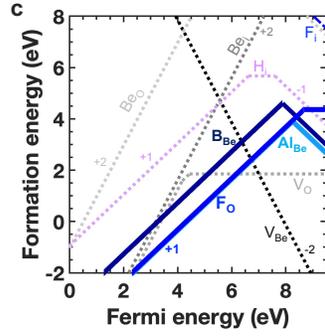
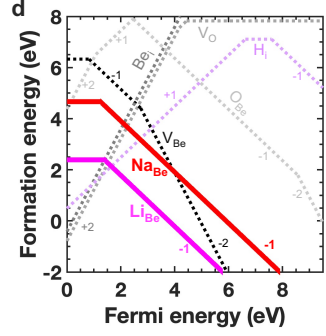
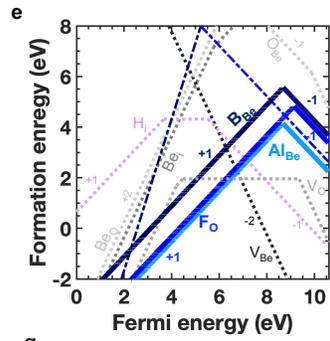
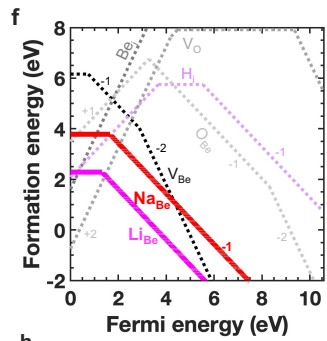
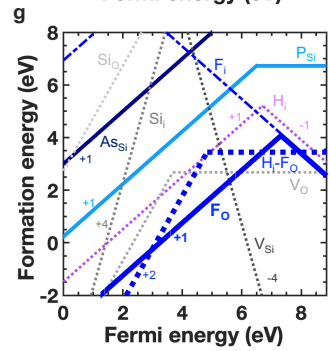
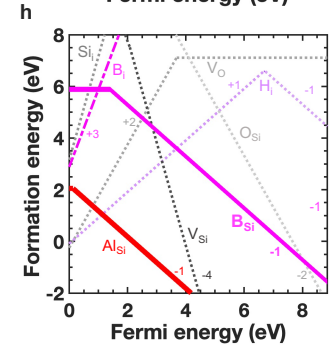
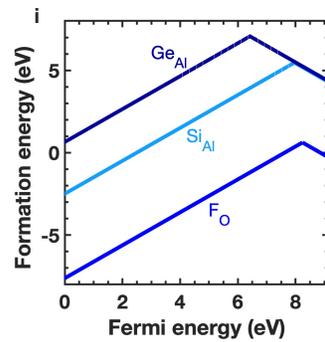
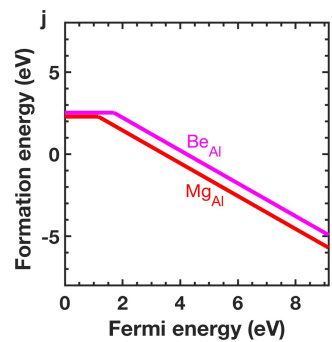

**Extended Data Fig. 4. First-principles calculation of defects and dopants. a, c, e, g, i,** Formation energy of donor impurities and intrinsic defects in **a,** rs-MgO, **c,** zb-BeO, **e,** wz-BeO, **g,** r-SiO$_2$, and **i,** Al$_2$O$_3$ as a function of the Fermi level under cation-rich/O-poor conditions. **b, d, f, h, j,** Formation energy of acceptor impurities and intrinsic defects in **b,** rs-MgO, **d,** zb-BeO, **f,** wz-BeO, **h,** r-SiO$_2$, and **j,** Al$_2$O$_3$ as a function of the Fermi level under cation-poor/O-rich conditions. **a-b,** For rs-MgO, F$_O$ and Al$_{Mg}$ are shallow donors with ionization energies less than 0.07 eV, while Li$_{Mg}$ and Na$_{Mg}$ are shallow acceptors with ionization energy of 0.07 eV and 0.17 eV, respectively. **c-f,** For BeO polytypes, the donors form DX centers with deep ionization energies larger than 1.2 eV, with the exception of F$_O$ in zb-BeO. Acceptors have deep ionization energies ranging from 1.2 eV to 2.0 eV. **g-h,** For r-SiO$_2$, the donors form DX centers with deep ionization energies of 1.53 eV for F$_O$, while Al$_{Si}$ is shallow acceptor with ionization energy of 0.13 eV while B$_{Si}$ has deep ionization energies of 1.41 eV. **i-j,** For Al$_2$O$_3$, the donors form DX centers with deep ionization energies of 0.763 eV for F$_O$, 1.18 eV for Si$_{Al}$, 2.82 eV for Ge$_{Al}$, while acceptors have deep ionization energies of 1.67 eV for Be$_{Al}$ and 1.17 eV for Mg$_{Al}$. Formation and ionization energies of intrinsic defects are reported in ref.[15]

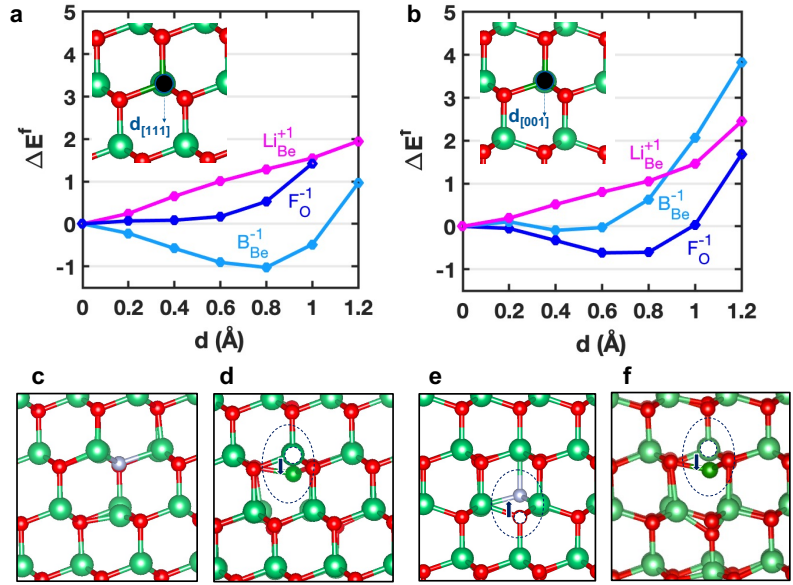

**Extended Data Fig. 5. Stability of DX-centers of donor and acceptor impurities in BeO polytypes. a-c,** Configuration coordinate diagram for the displacement of donor impurities with an extra electron (F$_O^{-1}$ and B$_{Be}^{-1}$) and acceptor impurities with an extra hole (Li$_{Be}^{+1}$) in **a,** zincblende- (zb-), and **b,** wurtzite (wz-) BeO. The formation energy of a defect ($\Delta E^f$) is calculated as a function of the displacement along the direction of an interstitial site ([111]-direction for zb, [001]-direction for wz) and is referenced to the energy of the configuration where the impurity atom is in the substitutional site. The DX-center is the most stable donors state, with the exception of F$_O$ in zb-BeO, while the AX-center is unstable for acceptors in both BeO polytypes. **c-f,** The defect configuration after final relaxation for F$_O^{+1}$ and B$_{Be}^{+1}$ in **c-d,** zb, and **e-f,** wz BeO.

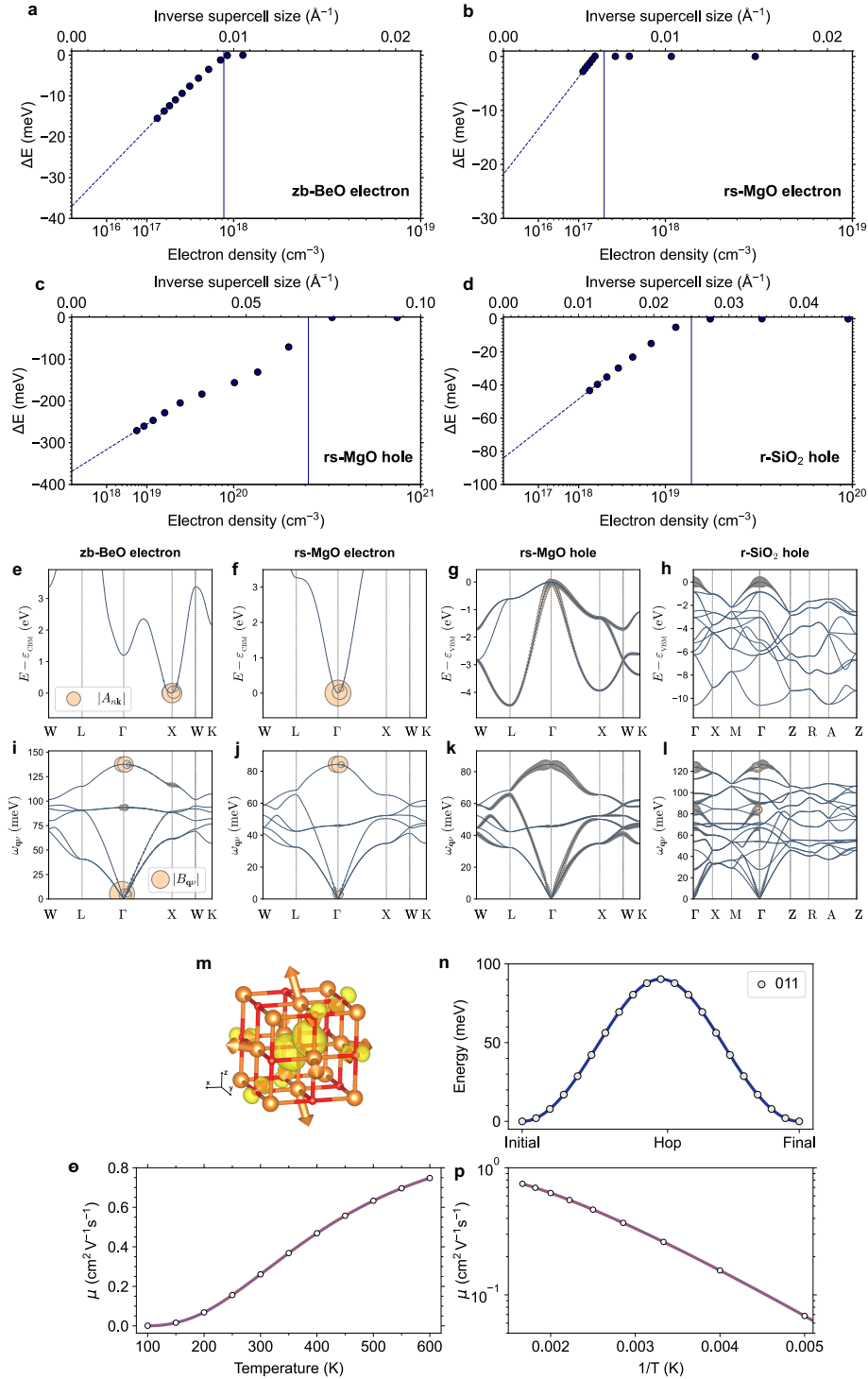

**Extended Data Fig. 6. First-principles calculations of polarons. a-d**, Formation energy of **a**, electron polaron in zb-BeO, **b**, electron polaron in rs-MgO, **c**, hole polaron in rs-MgO, and **d**, hole polaron in r-SiO$_2$ as a function of the supercell size, varied by increasing the Brillouin zone sampling to solve the *ab* initio polaron equations.[16] The zero of the energy axis is set to the energy of the undistorted structure. The vertical line marks the critical density for the metal-insulator transition. The dashed line represents the extrapolation of the polaron energy to the limit of infinite

supercell. **e-l**, Expansion coefficients for **e**, the polaron wave function in terms of Kohn-Sham orbitals and **i**, the atomic displacements in terms of phonon modes for the electron polaron in zb-BeO. Same as **e-i** in the case of **f,j**, electron polarons in rs-MgO, **g,k**, hole polarons in rs-MgO and **h,l**, hole polarons in r-SiO$_2$, respectively. The dominant contribution from the conduction/valence band edges and longitudinal-optical phonon modes indicate the formation of large Fröhlich-type electron polarons in zb-BeO, rs-MgO and hole polarons in r-SiO$_2$. The spread of the coefficients across the Brillouin zone in the case of hole polaron in rs-MgO indicates the formation of small polarons. **m**, Hole polaron in rs-MgO. Oxygen and magnesium atoms are represented by the red and orange spheres, respectively. The arrows represent the atomic displacements. The yellow isosurface represents the polaron charge density, at 1% of its maximum value. **n**, Energy barriers for polaron hopping to nearest neighbor oxygen site. The filled circles represent polaron energies for different configurations across the (011) direction. The intermediate configurations are obtained by performing a linear interpolation between the initial and final displacement configurations. Solid lines are guides to the eye. **o-p**, Temperature-activated mobility calculated from energy barrier in **n**, using the Emin-Holstein-Austin-Mott (EHAM) theory.[17]

**Extended Data References**